\documentclass[conference]{IEEEtran}
\IEEEoverridecommandlockouts
\usepackage{cite}
\usepackage{amsmath,amssymb,amsfonts}
\usepackage{algorithmic}
\usepackage{graphicx}
\usepackage{textcomp}
\usepackage{xcolor}
\usepackage{subfig}
\usepackage{dblfloatfix}

\def\BibTeX{{\rm B\kern-.05em{\sc i\kern-.025em b}\kern-.08em
    T\kern-.1667em\lower.7ex\hbox{E}\kern-.125emX}}
\begin{document}

\title{This Car is Mine!: Automobile Theft Countermeasure Leveraging Driver Identification with Generative Adversarial Networks}

\author{\IEEEauthorblockN{Kyung Ho Park}
\IEEEauthorblockA{\textit{Graduate School of Information Security} \\
\textit{Korea University}\\
Seoul, Republic of Korea \\
kyungho96@korea.ac.kr}
\and

\IEEEauthorblockN{Huy Kang Kim}
\IEEEauthorblockA{\textit{Graduate School of Information Security} \\
\textit{Korea University}\\
Seoul, Republic of Korea \\
cenda@korea.ac.kr}
}

\maketitle

\begin{abstract}
As a car becomes more connected, a countermeasure against automobile theft has become a significant task in the real world. To respond to automobile theft, data mining, biometrics, and additional authentication methods are proposed. Among current countermeasures, data mining method is one of the efficient ways to capture the owner driver's unique characteristics. To identify the owner driver from thieves, previous works applied various algorithms toward driving data.

Such data mining methods utilized supervised learning, thus required labeled data set. However, it is unrealistic to gather and apply the thief's driving pattern. To overcome this problem, we propose driver identification method with GAN. GAN has merit to build identification model by learning the owner driver's data only. We trained GAN only with owner driver's data and used trained discriminator to identify the owner driver. From actual driving data, we evaluated our identification model recognizes the owner driver well. By ensembling various driver authentication methods with the proposed model, we expect industry can develop automobile theft countermeasures available in the real world.
\end{abstract}


\section{Introduction}
Along with the development of technology, the number of computing devices inside a vehicle increased. As representatives, Controller Area Network (CAN) and Electronic Control Units (ECUs) enable communication within a car, In-vehicle Infotainment (IVI) lets the driver to enjoy various contents, and sensors such as camera, LiDAR, and RADAR make the driving environment safer. Although the purpose of computing devices varies, the entire car system is oriented to enhance communication around the vehicle. On the other hand, threats emerged in communication-related aspects accompanying automobile theft. Following past researches, more communication around the car generates more threats to the automobile system \cite{checkoway2011comprehensive}. 


Nowadays, enormous investment and research effort started to focus on Vehicle-to-Everything (V2X) communication and autonomous driving, thus cars become more connected. These approaches promote benefits but create security concerns related to automobile theft at the same time. If past car thieves physically broke windows for theft, modern thieves would perform cyber attacks toward the car. In July 2015, Reuters reported a former U.S National Security Agency hacker Charlie Miller and IOActive researcher Chris Valasek used a feature in the Fiat Chrysler telematics system to break into the car running on a highway. Miller and Valasek turned on the Jeep Cherokee’s radio and activated other inessential devices \cite{reuter}. In the era of V2X communication and autonomous driving, automobile thefts do not simply occur by physical access, but also occur from threats at cyber world. Therefore, automobile theft countermeasure at both real world and cyber world is highly required.

Recently, several countermeasures against automobile theft were proposed. For example, researchers suggested biometrics methods based on image processing with face, fingerprint, and iris or signal processing method for speech recognition \cite{villa2018survey}. However, car manufacturers had to jump over hurdles for implementation: hardness of achieving high accuracy, inconsistent performance, inconvenience of installing additional hardware, which results in inefficiency \cite{nimbhorkar2015survey}. To overcome shortcomings of the biometrics, data mining method appeared on the stage. Data mining method captures the pattern of the owner driver, and compare newly recognized data with the owner driver's data. When people drive through the road, all drivers have different driving styles. For example, some drivers accelerate as fast as they can while others smoothly accelerate. Some drivers push the brake pedal multiple times, while others make a sudden stop. By utilizing these different driving patterns, numerous researches introduced data mining methods for driver identification.

However, past data mining methods have a limit to be utilized as automobile theft countermeasure. Previous data mining methods require well-labeled data set for the owner driver and thieves. But it is hard to collect or even predict automobile thief data; thus, it is impossible to train the model. Moreover, collecting data from a car becomes a resource-consuming task when it goes on a large scale, and the privacy issue exists in the data collection process. Data related to driving habits can be classified as private information, and car manufacturers might have a hurdle to collect for training \cite{cpo}.

In this work, we developed automobile theft countermeasure with driver identification, which necessitates the owner driver's data only. To hedge data dependency of previous identification models, we leveraged Generative Adversarial Networks (GAN) which only requires the owner driver's data for training. GAN is composed of a generator creating fake output, and discriminator classifying whether input data is real or fake. GAN is commonly used to create artificial data with the generator, but we used discriminator to classify whether provided driving data is owner driver's data (real) or not (fake). We trained and evaluated GAN-based identification model with test data set composed of four different drivers. Following this research, our contributions are as follows:

1) Practicality: Our model does not require thief data at the training stage. Thus, it can be applied as an automobile theft detection method in the real world.

2) Economic Feasibility: As our model analyzes data extracted from CAN, we can simply extract data from the vehicle without resource-consuming hardware installation.

3) Privacy: We designed a model which does not require non-owner driver's data for training; therefore, car manufacturers can hedge the risk of privacy issue for commercialization.

4) Security: As our model is built by GAN, a black-box model without using pre-defined ruleset; thus, malicious intruders cannot figure out the driver identification logic easily. 

\section{Literature Review}
Among numerous driver identification methods, data mining method showed precise identification performance rather than others. Following target data used in the analysis, previous data mining methods are categorized as Table 1.

\begin{table*}[!t]
\centering
\caption{Researches on data mining methods for driver identification}\label{tab1}
\resizebox{0.8\textwidth}{!}{
\begin{tabular}{|c|c|c|c|}

\hline
Target Data & Specific Data Type & Identification Methodology & Reference\\

\hline
\centering
Simulated Data & Behavior Signals & Statistical Algorithm & \cite{wakita2006driver}\\
& Accelerator, Steering Wheels & Statistical Algorithm & \cite{zhang2014study}\\

\hline
Mixed Data & Behavior Signals & Statistical Algorithm & \cite{miyajima2007driver}\\
& Vehicle Sensors and & Statistical Algorithm and & \cite{wahab2009driving}\\
& Speech, Image, Location Signals & Neural Networks & \\

\hline
Sensor Data & Embedded Physical Sensor & Statistical Algorithm & \cite{nishiwaki2007driver} \\
& CAN & Statistical Algorithm & \cite{choi2007analysis} \\
& CAN & Machine Learning Algorithm & \cite{enev2016automobile} \\
& CAN & Neural Network & \cite{zhang2019deep}, \cite{kwak2016know} \\
\hline
\end{tabular}
}
\end{table*}

Data mining method started with simulated driving data. T. Wakita \textit{et al.} proposed driver identification method analyzing observed behavior signals when a driver follows another car. They collected data from driving simulator composed of the steering wheel, pedals, and LCD monitor showing driving scene. They identified multiple drivers with Gaussian Mixture Model (GMM) \cite{wakita2006driver}. Zhang \textit{et al.} proposed a driver identification model with Hidden Markov Model (HMM) analyzing accelerator and steering wheel data. \cite{zhang2014study}. Above researches showed data mining method with driving data can identify drivers. But there exists a limitation that the simulated environment is hard to reflect noises from the real world.

To reflect actual driving status, some researches proposed driver identification model with mixed data from the real world and simulator. Nishiwaki \textit{et al.} collected data from embedded sensors in a car, and built a driver identification model based on driving characteristics illustrated in spectral features. They focused behavioral signals such as brake pedal pressures. Training data was collected by 276 drivers, using a specially designed vehicle. Identification model with GMM resulted in 76.8\% accuracy, showing the possibility of driver identification with actual sensor data \cite{nishiwaki2007driver}. Although data mining with sensor data established improvement, data collection process such as device installation was resource-consumable. At the industry level, it is not economic to bare cost for data collection devices. Instead of resource-consuming physical sensor data, researchers started to focus on CAN data. By embedding cables to a car, it was much easier to collect driving data from CAN.

Researchers actively utilized the convenience of CAN data. Choi \textit{et al.} collected CAN data from 9 drivers, applied statistical methods on driver identification \cite{choi2007analysis}. Although accuracy was 25\%, they showed CAN data analysis is a feasible method with the lightweight data collection process. Enev \textit{et al.} collected CAN data at the parking lot and city road from 15 different drivers, and achieved 100\% accuracy with ensembling machine learning algorithms \cite{enev2016automobile}. Zhang \textit{et al.} proposed identification methods leveraging Deep Neural Networks (DNNs) \cite{zhang2019deep}. They experimented various mixtures of neural networks composed of Convolutional Neural Network (CNN) and Recurrent Networks such as LSTM (Long Short Term Memory). From previous works, we recognized CAN data contain meaningful features, and data mining method is effective for driver identification.

In this work, we present the improved data mining method for automobile theft countermeasure. Previous methods precisely identified many drivers, but it is only meaningful when every driver's data are trained. If unlabeled data is provided, the model gets confused as it counters a new pattern which does not exist in trained history. When we design automobile theft countermeasure, it is impossible to collect and train thief data to the model. Thus, the identification model should classify unlabeled data as a thief without training. Then, the model detects unknown patterns as thieves, and recognize the trained pattern as the owner driver. We acknowledge previous works are significant at the labeled data set, such as co-owners of a family car. But they are inadequate to be applied at automobile theft with unlabeled data. To overcome this limitation, we suggest the driver identification model only with owner driver's data. To achieve this goal, Kang \textit{et al.} previously suggested driver identification method utilizing clustering algorithms \cite{kang2019automobile}. In this work, we leveraged RGAN to provide a baseline model of automobile theft countermeasure which is applicable in the real world.

\section{Proposed Methodology}
\subsection{Data Collection}
We accumulated driving data from four different drivers with a single vehicle, Hyundai YF Sonata to minimize noise from vehicle difference. While a driver is driving along the route illustrated in Fig. 1, we collected CAN data utilizing On Board Diagnostic 2 (OBD-II) and CarbigsP as scanning tool to extract data. Leveraging that OBD-II diagnoses status of ECUs and return specific value of them, we accumulated driving data with 51 features, recorded per second. At experimental driving, four drivers drove along the route following traffic rules of South Korea.


\begin{figure}[!h]
\centering
\includegraphics[width=0.38\textwidth]{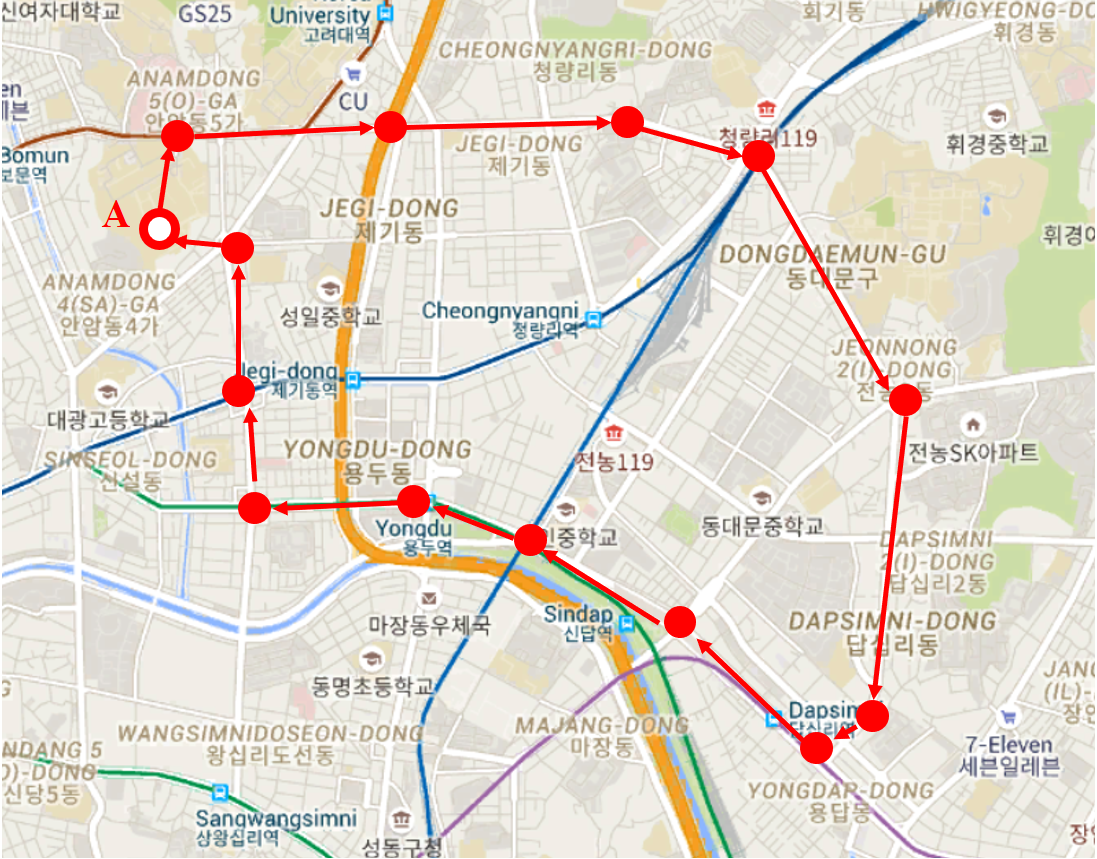}
\caption{Driving route for data collection}
\end{figure}

\begin{figure}[!h]
\centering
\includegraphics[width=0.38\textwidth]{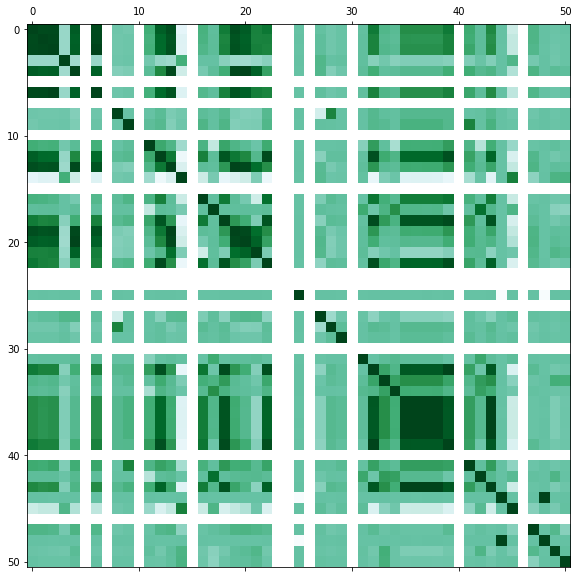}
\caption{Visualized correlation matrix composed of 51 features }
\end{figure}

\subsection{Feature Engineering}
Feature engineering process eliminates potential defects at model training. For better performance, we managed three levels of feature engineering process: Exclusion of highly-correlated features, filtering non-influential features, and sliding window with normalization.

\subsubsection{Exclusion of Highly-Correlated Features}
Raw driving data contains 51 features, and some of these features have a comparatively high level of correlation. When highly-correlated features are provided into the model, they blur unique characteristics of owner driver’s driving pattern. Following experimental trials, we analyzed a pair of features with the absolute value of correlation higher than 0.95 are highly-correlated. Fig. 2 visualizes correlation matrix of 51 features selected from four owner drivers' single trip data. The darker color shows a high level of correlation, and lighter color shows a low level of correlation. We excluded highly correlated features and resulted in 32 essential features as Table 2.

\begin{table*}
\caption{Essential features from CAN bus data}\label{tab1}
\centering
\resizebox{0.75\textwidth}{!}{%

\begin{tabular}{|l|l|}
\hline
Category & Features\\
\hline
Fuel & Fuel consumption, Short term fuel trim bank, Long term fuel trim bank,\\
& Intake air pressure, Engine fuel cut off, Decreased fuel cut off\\
\hline
Engine & Engine speed, Current spark timing, Engine coolant temperature,\\
& Target engine idle speed, Maximum engine torque, Minimum engine torque,\\
& Calculated LOAD value, Standard torque ratio, Engine torque cutoff,\\
& Engine speed increase\\
\hline
Transmission & Friction torque, Torque converter speed, Current gear level,\\
& Transmission oil temperature, Torque converter turbine speed, Converter clutch,\\
&Gear choice, Steering wheel speed, Steering wheel angle,\\
&Front-left wheel speed, Front-right wheel speed,\\
&Rear-left wheel speed, Rear-right wheel speed\\
\hline
Misc. & Car speed, Brake switch, Road gradient\\
\hline
\end{tabular}%
}
\end{table*}

\subsubsection{Filtering Non-Influential Features}
Non-influential features also blur identification model. As model interprets distinguishable features through training, excluding features without significant difference enhances model performance. We excluded features fulfilling one of any statistical rules presented below:

\begin{itemize}
    \item \textit{Rule 1) Missing Value}: Certain feature contains null value throughout the driving.
    \item \textit{Rule 2) Feature Indifference}: A value of certain feature is indifferent at each driver
    \item \textit{Rule 3) Feature Invariance}: Aggregated value of certain feature is zero and the standard deviation of that feature is also zero at each driver.
\end{itemize}

\noindent Features satisfying Rule 1) implicate the existence of error during the data extraction process, thus we dropped them. Rule 2) proposes the feature does not imply any distinct characteristics among drivers. As indifferent features can blur the detection model, we also eliminated features fulfilling Rule 2). Lastly, Rule 3) implies the existence of data extraction error. If an unknown error happens during the data extraction process, we checked OBD-II records consistent zero value. We dropped features fulfilling the third rule to minimize data extraction error.

\subsubsection{Sliding Window with Normalization}

To transform raw data into the trainable form, we performed window sliding. We set 33 seconds as the fixed length of time window. By sliding time window through a trip time, we created multiple training data. We normalized all training data employing the following equation below, letting different scales of features ranged in a specific boundary.
\[ X_{normalized} = \frac{X_i - Min(X_i)}{Max(X_i) - Min(X_i)}\]\

\subsection{Algorithm: Recurrent Generative Adversarial Network}
We set GAN as classification algorithm. GAN is deep neural networks comprised of two objects: Generator and Discriminator. The generator creates realistic data to deceive discriminator, while the discriminator classifies whether input data is real or fake. By composing original GAN, numerous derived GAN models have been proposed following which type of data to train \cite{esteban2017real}. Acknowledging driving data has temporal dynamics as time-series movement, we used Recurrent Generative Adversarial Networks (RGAN) which generator and discriminator are composed in Recurrent Neural Networks (RNN) architecture. As numerous works have proved RNN performs well with time-series movement, we expect RGAN learns driving data effectively.

\begin{figure*}[t!] 
\centering
\includegraphics[width=0.8\textwidth]{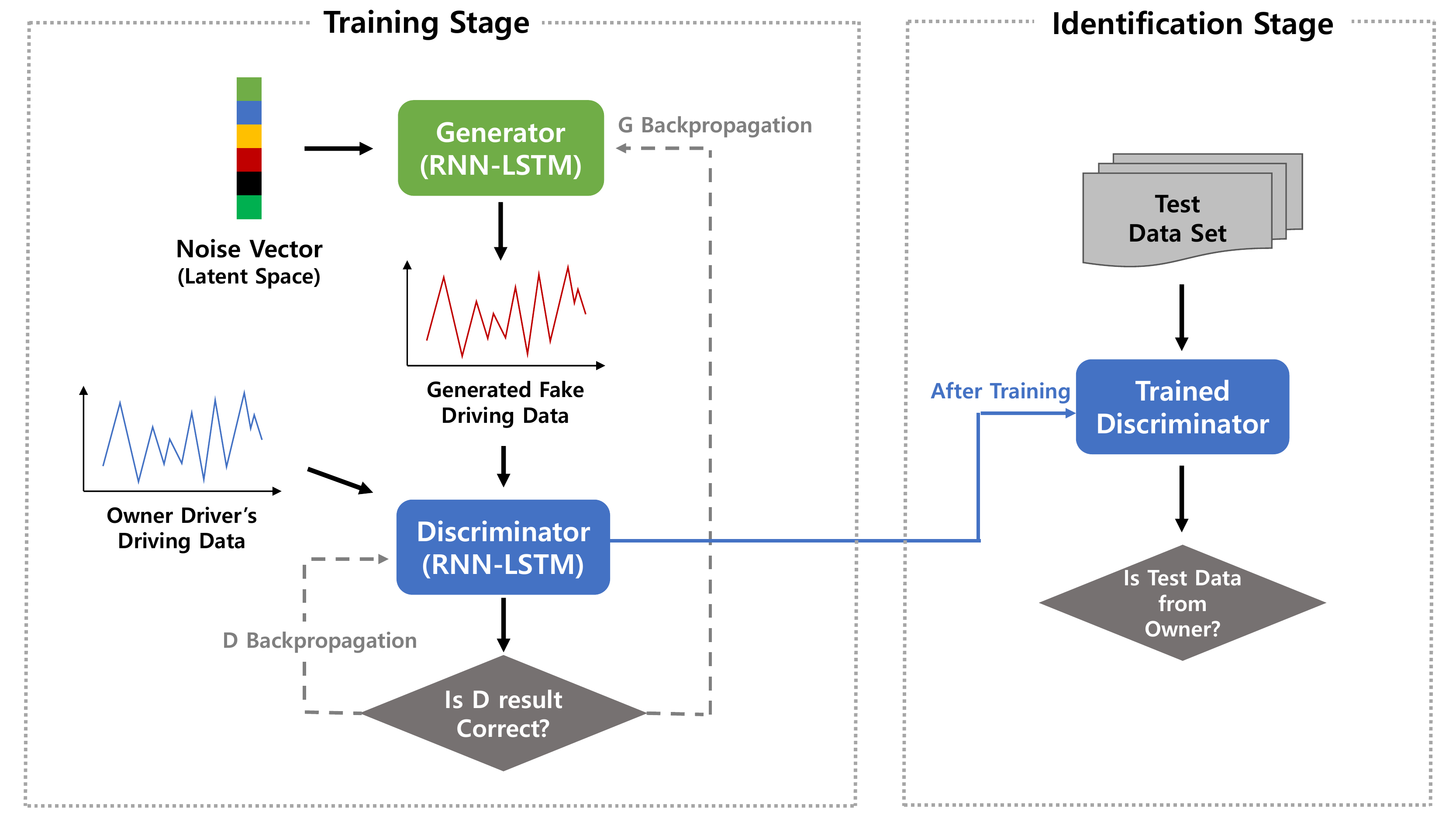}
\caption{Driver identification model with RGAN} \label{fig3}
\end{figure*}

As illustrated in Fig. 3, the identification model consists of RNN-LSTM to hedge vanishing gradient problem of RNN. Generator receives randomly distributed noise and return fake time-series driving data. Discriminator receives both real training data and generated fake data to compute probability whether data is real or not. Through experimental trials, we optimized discriminator and generator with Gradient descent optimizer and Adam optimizer, respectively. We also observed the generator requires more iteration of training rather than discriminator, consuming more effort to the deceive discriminator. We trained the owner driver's data and evaluated the performance of trained discriminator at the test set.

\section{Experiment Result}
\subsection{Experiment setup}
We performed an experiment with four different drivers labeled as driver A, B, C, D, and picked one driver as owner and others as thieves. If driver A is selected as the owner, B, C and D are labeled as thieves. The training set is only composed of owner driver's driving data, and test set consists of both owner driver and thief drivers. We formally expected automobile theft rarely occurs during the car's life cycle, thus composed test set following ratio of 8:2 for owner and thief drivers respectively. The average time length of the training set is 1986 seconds. Thus, the model learned owner driver's pattern during 33 minutes of driving.

\subsection{Evaluation}
We set a goal of proposed model as detecting the occurrence of automobile theft. Thus, we defined false positive error and the false negative error as below:
\begin{itemize}
    \item \textit{False Positive Error}: Identify thieves as the owner driver
    \item \textit{False Negative Error}: Identify the owner driver as thieves
\end{itemize}

We evaluated detection performance with four metrics: accuracy, precision, recall, and F1 score. Accuracy is a ratio of correctly identified result to the total identification results, and it provides the most intuitive performance evaluation. As test sets are not distributed in symmetric way, we considered other evaluation criteria for accurate evaluation. Precision is the ratio of correct identification results to the total identification results. High precision shows the low false positive error, implicating thieves are not identified as the owner driver. Recall is the ratio of correct identification to all of actual labels. High recall implicates the low false negative error that owner drivers are not identified as thieves. Lastly, F1 score is the weighted average value of precision and recall, considering both false positive error and false negative error. Following four evaluation metrics, Table 3 shows the experiment result.

Trained model recognized the owner driver well with reasonable accuracy, and the result shows lower recall compared to precision. In the viewpoint of security and safety, we allowed this low recall. Both false positive error and false negative error at theft detection are critical to driver experience. Low precision with high false positive error can cause a security problem, which brings direct loss of asset. Low recall with high false negative error generates a usability problem, that the owner driver gets annoyed from false theft detection. In this research, we prioritized security rather than usability. We analyzed the first and foremost goal of theft countermeasure is securing a valuable asset, the car. As false positive errors are hard to be mitigated, thus we decided proposed models should have high precision to reduce serious damage toward the asset. On the other hand, other authentication modules can supplement to recognize misplaced owner driver, then false negative errors are relatively tolerable. To hedge the fatal situation from such security problems, we designed the model to minimize false positive error. We evaluate the proposed model suggests a stable baseline for owner driver identification only with owner driver's data. From this baseline, applying supplementary methods discussed in the Discussion section would create theft countermeasure with reduced usability problem.

\begin{table}
\caption{Driver identification experiment result}\label{tab1}

\centering

\begin{tabular}{|c|c|c|c|c|c|}

\hline
Driver & Training set & Accuracy & Precision & Recall & F1 Score\\
\hline
A & 1930s & 0.891 & 0.913 & 0.738 & 0.788\\
B & 2034s & 0.862 & 0.796 & 0.741 & 0.763\\
C & 2002s & 0.892 & 0.94 & 0.73 & 0.783\\
D & 1978s & 0.894 & 0.846 & 0.808 & 0.824\\
\hline
Average & 1986s & 0.884 & 0.873 & 0.754 & 0.789\\
\hline

\end{tabular}
\end{table}

\section{Use Case Scenario}

\subsubsection{Implementation} Our detection method can be developed as an independent application on automotive operating systems. As a benchmark case, previous research suggested driver identification application on Automotive Grade Linux \cite{el2019improving}. Similarly, we expect our theft detection application can be implemented on IVI, and provide User Interface (UI) to the owner driver. This application at IVI enables the owner driver to start training stage of theft detection model. Under authenticated circumstances such as right after the purchase of the new car, the owner driver can initiate data collection to create training set.

\subsubsection{Model training} Leveraging that OBD-II can send CAN data via Bluetooth communication, our application would collect CAN data easily. Following the experiment setup, our model necessitates at least 30 minutes of driving data as a training set. From training data collected from OBD-II, our application learns the driving pattern of the owner driver. If computing power at the given automotive environment is not enough, our application can send training set to the independent server through telematics channel. Then, the server with enough computing power trains detection model, and send the trained model back to the vehicle.

\subsubsection{Real-time detection} After the training stage, the trained model performs real-time theft detection for every validation data created through the driving. The proposed method utilizes the fixed time window of 33 seconds; thus our model performs real-time detection for validation data with 33 seconds of length. When theft is detected, the application sends a notification to the owner driver through text message to the owner driver. If the owner driver answers that theft is actually happened, application notifies the occurrence of theft to appropriate authorities such as police or car manufacturers. In case of urgent situations, the application would let Advanced Driver Assistance System (ADAS) to lock the car door and ring theft alarms. If the owner driver answers that theft detection was wrong, application accumulates these wrong detection results as an additional training set, and re-train the model for sustainable maintenance of detection model.

\subsubsection{Comparative advantage} The key advantage of our baseline model is generality, that our detection model can be easily applied to common cars. Previous works showed the possibility of theft detection leveraging various devices. For example, car manufacturers would utilize in-car cameras at theft detection, which was originally designed to check whether the driver is paying attention. However, past methods necessitate additional sensors such as a camera. Therefore, they cannot be applied to old cars which are not compatible with additional modules. Moreover, the installation of additional sensor creates economic burden toward car manufacturers. Compared to previous methods, our theft detection model can be applied to common cars as CAN is boarded in most cars. Furthermore, our model can be implemented without installing additional sensors. Thus, car manufactures do not have much burden of installation cost. Based on these key advantages, we analyzed our proposed method can be improved with further research effort, which is elaborated in the following section.

\section{Discussion}

\subsubsection{Improving detection performance with ensembling} Ensembling different method would provide more accurate theft detection performance. As both false positive error and false negative error of theft detection should be mitigated, additional authentication methods would supplement proposed baseline model. For instance, device pairing authentication method would mingle well with proposed model by reducing false negative error. Embedded pairing module authenticates driver when the registered device is paired. As people carry electronic devices such as smartphone, pairing information is highly applicable in daily circumstances. If the module is paired, our model can adjust the identification result reflecting result of device authentication. With collaborative driver identification, our model can provide enhanced performance enough to be used in the real world.


\subsubsection{Theft detection at various driving circumstances} Our proposed model can be enhanced by considering various circumstances. First, our model shall be trained to recognize multiple co-owners for a single car. Our experiment assumed a case that there is a single owner driver for a single car. However, there frequently exists multiple co-owners in a real world. In the case of a family car, our detection model should not identify family members as thieves. Therefore, it is necessary to enhance our detection method to recognize multiple legitimate owner drivers. Furthermore, the proposed method can be validated at different road conditions such as highways or parking lots. Lastly, for the better applicability, we will improve our proposed method to use smaller size of training data.  


\section{Conclusion}
In this paper, we propose automobile theft countermeasure utilizing driver identification method. Past data mining methods identified driving patterns well, but there existed an obstacle that it only identifies labeled data. Due to the hardness of collecting thief data, identification model should identify the owner driver well with unlabeled data. To hurdle over this problem, we employ RGAN which only requires the owner driver's data. The proposed model learned owner driver's driving characteristics, and recognized the owner driver from thieves with reasonable performance.

Our automobile theft countermeasure suggests meaningful contributions. First and foremost, the model only requires the owner driver's data and classifies any abnormal driving pattern as a thief. Our model analyzes CAN data, thus accompanies lightweight data collection process. As it only necessitates the owner's data, we can also hedge data privacy problem. Finally, our countermeasure is hard to be neutralized as RGAN is a black-box model. We expect collaborative identification with supplementary authentication methods enhances result precisely. In pursuit of precise automobile theft countermeasure, we will further research on driver identification at various driving circumstances with RGAN.

\section*{Acknowledgements}
This work was supported by Samsung Research Funding \& Incubation Center for Future Technology under Project Number SRFC-TB1403-51.

\bibliographystyle{IEEEtran}
\bibliography{reference}

\end{document}